# The Effect of Dipole from $\gamma-$AgI Substrates on Heterogeneous Ice Nucleation

Hao Lu, Quanming Xu, Chaohong Wang, Jianyang Wu[*], Rongdun Hong[*],

Xiang-Yang Liu, Zhisen Zhang[*]

**Department of Physics, Research Institute for Biomimetics and Soft Matter, Fujian Provincial Key Laboratory for Soft Functional Materials Research, Jiujiang Research Institute, Xiamen University, Xiamen, Fujian 361005, China**

Email: zhangzs@xmu.edu.cn, jianyang@xmu.edu.cn, rdhong@xmu.edu.cn

**ABSTRACT:** Heterogeneous ice nucleation is one of the most common and important process in the physical environment. AgI has been proved to be an effective ice nucleating agent in the process of ice nucleation. However, the microscopic mechanism of AgI in heterogeneous ice nucleation has not been fully understood. Molecular dynamics simulations are applied to investigate the ability of which kinds of $\gamma$-AgI substrate can promote ice nucleation by changing the dipole of $\gamma$-AgI on the substrate, we conclude that the dipole of $\gamma$-AgI on the substrate can affect the conformation of ice nucleation. The surface ions with positive charge on the substrate may promote ice nucleation, while there is no ice nucleation founded on the surface ions with negative charge. $\gamma$-AgI substrates affect ice nucleation through adjust the orientations of water molecules near the surfaces.

## Introduction

Ice nucleation is important in many ways, such as climate, microbiology and atmospheric science.[1-3] There are two types of ice nucleation, one is homogeneous ice nucleation and the other is heterogeneous ice nucleation.[4-6] Homogeneous ice nucleation does not require an external medium, but a lower environemnt temperature is needed, homogeneous ice nucleation would not occur when the temperature is higher than 235 K (about -38 °C)[7-10]. Heterogeneous ice nucleation can occur at a warmer temperature, but heterogeneous ice nucleation requires the participation of external medium. This external medium is called ice nuclei (IN), and many substrates have been proven to be effective heterogeneous IN, such as dust,[11] graphitic,[12-13] silver iodide,[1-2, 14-16] kaolinite,[17-19] etc.

Silver iodide can promote the temperature required for water freezing from -38 ℃ to -3 °C,[20] which has been regarded as one of the best heterogeneous nucleation IN known and has been widely used in cloud seeding. Silver iodide contains two polymorphs under atmospheric conditions, $\gamma$ - AgI and $\beta$ - AgI, $\gamma$ - AgI is a kind of cubic[21-22] crystal and $\beta$ - AgI is hexagonal, both have been concerned in recent years.[4, 16, 23-26] Good lattice matching between AgI and hexagonal ice is considered to be the main reason for promoting ice nucleation in traditional views,[16, 20] but recent research shows that lattice match is not the only factor in promoting ice nucleation.[1-2, 27-29] Some researches about ice nucleation on $\gamma$ - AgI surface have illustrated that ice nucleation occurs when silver ions are exposed on the $\gamma$ - AgI surface with charge ranging from 0.2e to 0.6e, rather than iodine ions exposed on the surface of $\gamma$ - AgI.[4] They came to a conclusion that there are no ice nucleation found on iodine ions exposed $\gamma$ - AgI surface for the structure formed by the water molecules on the surface does not match the structure of the hexagonal ice. However, the explanation of these phenomena has not been mentioned yet.

In our work, two kind of substrates, one with silver ions exposed on $\gamma$ - AgI surface and one with iodine ions exposed on $\gamma$ - AgI surface are created. Changing the position of silver ions and iodine ions only without changing the lattice matching of the $\gamma$ - AgI structure. Using molecular dynamics (MD) simulation to analyze the ice nucleation ability of $\gamma$ - AgI by changing the charge of the exposed surface ions, calculate the time required for ice nucleation in different systems and observe the condition of water molecules in these simulations. Our results show that the dipole of $\gamma$ - AgI on the substrate has a great influence in the process of ice nucleation. In these systems, ice nucleation is a combination of lattice matching and dipole of $\gamma$ - AgI on the surface. It is expected that the microscopic knowledge reported in our simulations can be used in designing the specific substrates which can promote or hinder ice nucleation.

## Simulation Details

All MD simulations were carried out under NVT condition employing the GROMACS 4.6.7 package.[30] LINCS algorithm was employed to restrict hydrogen bond lengths and angles in the programs.[31] The particle mesh Ewald method was used to calculate the electrostatic forces and Nose-Hoover thermostat was used to keep the temperature.[32-33] The TIP4P/Ice[34] water model (melting point 270 ± 3 K) was used in these simulations.

$\gamma$-AgI slabs were positioned on mirrored position in the simulations to eliminate long-range electric field produced by lattice truncation.[2] Four $\gamma$-AgI layers made up the $\gamma$-AgI slab each side and the AgI slabs were rigid during the simulations. The box's dimension was 5.197 × 5.197 × 8.000 nm$^3$, 4032 water molecules were contained in these simulation systems. Water molecules were closer to the silver ions in the model silver exposed $\gamma$-AgI (001) surface, and water molecules were closer to the iodine ions in the model iodide exposed $\gamma$-AgI (001) surface. The distance between each layer was twice as many as the distance between silver and iodine ions in vertical dimension on each layer, which is 0.325 nm.

A time step of 2 fs was employed and the MD lasted 200 ns in total. The configurations were saved every 10 ps during the simulations. A criteria was enacted that if the system can nucleate ice in 200 ns, the nucleation ability of this system will be admitted, otherwise not, even some systems can nucleate ice at a longer time scale. The temperature in the simulations was carried at 300 K for 1 ns and then decreased to 250 K, 20 K below the TIP4P/Ice water model melting point for the rest of the simulation time. The charge of each ion exposed on the surface ranging from -1.0e to 1.0e. The simulation results show that silver exposed $\gamma$-AgI (001) surface can nucleate ice with charge from 0.2e to 0.6e, while iodine ions exposed $\gamma$-AgI (001) surface can nucleate ice with charge from 0.2e to 0.5e.

## Results and Discussion

The two models silver ions exposed on γ - AgI (001) surface and iodine ions exposed on γ - AgI (001) surface results in different phenomenons which shown in Figure 1.

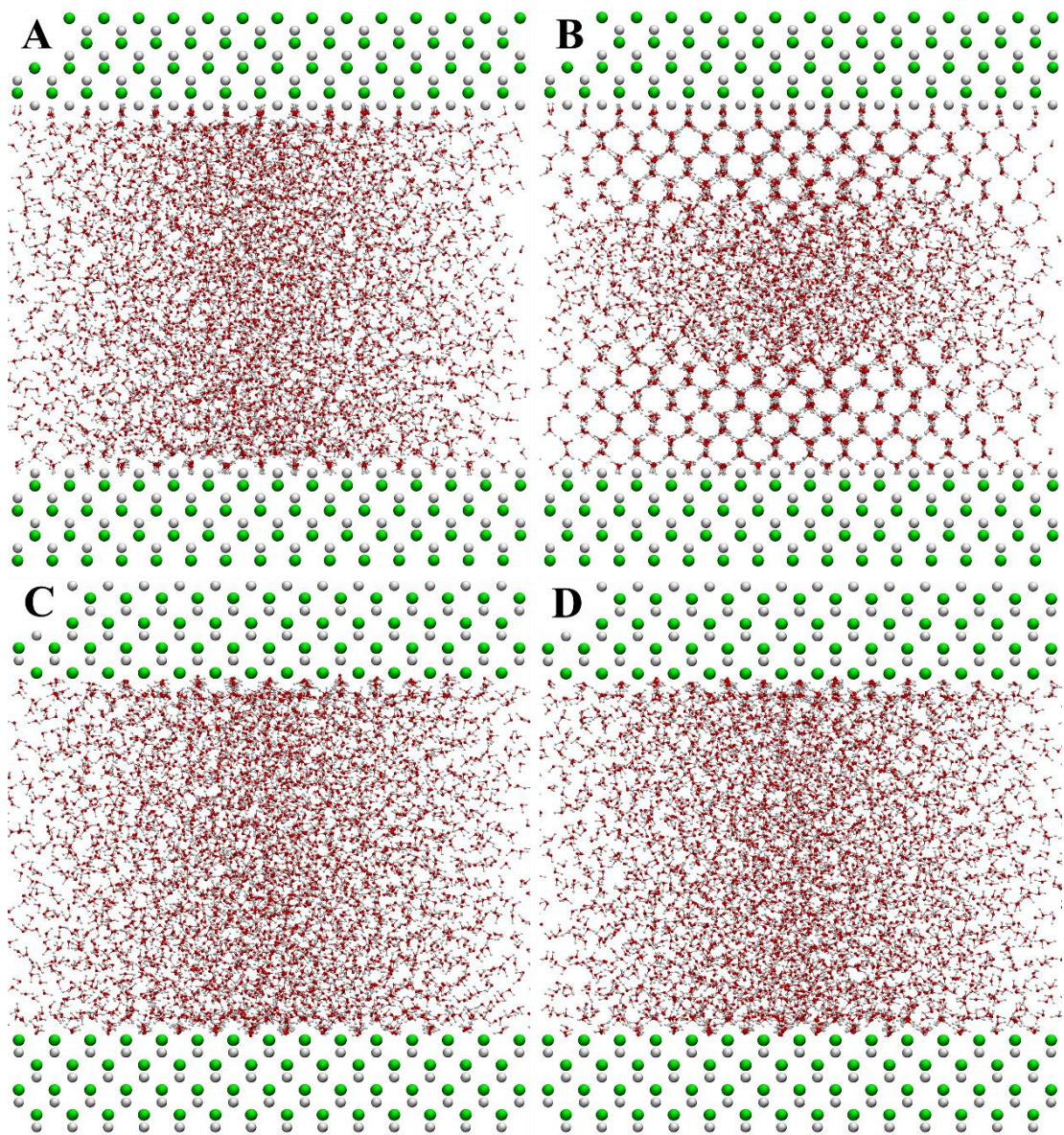

Figure 1. Snapshots of simulations on silver and iodine ions exposed γ-AgI (001) surface. Panels A and B are silver ions exposed as the outermost layers at 0 ns, 15 ns, respectively. Panels C and D are iodine ions exposed as the outermost layers at 0 ns, 200 ns, respectively. Silver ions, iodine ions, hydrogen atoms and oxygen atoms are silver, green, white and red, respectively. The charge of silver ions and iodine ions are 0.4e and -0.4e, respectively. The snapshots are rotated to the direction of (110) plane.

Figure 1 shows the snapshots of simulations at different times on silver exposed γ-AgI (001) surface (Figure 1A, 1B) and iodide exposed γ-AgI (001) surface (Figure 1C, 1D). The water molecules in the model silver exposed $\gamma$- AgI (001) surface take a short time scale from intial liquid state (Figure 1A) to ice state (Figure 1B), the water molecules in the model iodide exposed $\gamma$- AgI (001) surface still maintain a liquid state at 200 ns (Figure 1D), just the same as the initial liquid state (Figure 1C). The only difference between the two systems is the dipole direction of the ions on the substrates. To better understand the nucleation ability of different substrates, several simulations were carried out after adjusted the magnitude and orientation of dipole on the substrates to explore how do dipoles on the substrate affect ice nucleation. In the systems silver and iodine ions exposed $\gamma$- AgI (001) surface, the charge of ions exposed on the surface are changed from -1.0e to 1.0e with an interval of 0.1e, every simulation was carried out for 10 times. Simulation results are shown in Table 1. It is worth noting that some water molecules remaining exist in the system though most of them have grown into ice bulk (Figure 1B), this may be because the space between the substrates is not an integer multiple of the ice lattice size, the structure of the two subatrates are not strictly symmetric may be another reason.

Table 1: Freezing condition at different charges on silver and iodine ions exposed $\gamma$- AgI (001) surface.

| Charge | 0.1e | 0.2e | 0.3e | 0.4e | 0.5e | 0.6e | 0.7e | 0.8e | 0.9e | 1.0e |
|---|---|---|---|---|---|---|---|---|---|---|
| Ag—Positive | - | √ | √ | √ | √ | √ | - | - | - | - |
| Ag—Negative | - | - | - | - | - | - | - | - | - | - |
| I—Positive | - | √ | √ | √ | √ | - | - | - | - | - |
| I—Negative | - | - | - | - | - | - | - | - | - | - |

Table 1 shows the freezing condition at different charges on silver and iodine ions exposed $\gamma$- AgI (001) surface. The silver exposed $\gamma$- AgI (001) surface can nucleate ice with the charge of silver ions ranges from 0.2e to 0.6e, which is corresponding to the work of Stephen and Allen,[4] while the iodide exposed $\gamma$- AgI (001) surface can nucleate ice with the charge of iodine ions ranges from 0.2e to 0.5e, the detailed freezing time are provided in the Supporting Information S1 (Table S1). The freezing time of silver ions exposed $\gamma$- AgI (001) surface is a little shorter than iodine ions exposed $\gamma$- AgI (001) surface, especially when the charge of ions equals to 0.2e and 0.5e, it may be the difference in the radius of sliver ions and iodine ions leads to such a phenomenon.

A series of simulations were carried out which shown in Table 1, two representative systems are chosen to describe the principles. Figure 2 shows the snapshots of simulations on silver and iodine ions exposed $\gamma$ - AgI (001) surface, the charge of silver ions is -0.4e, and iodine ions is 0.4e. Compared with Figure 1 A-B, when change the charge of silver ions from 0.4e, to -0.4e, the system could not nucleate ice even in 200ns. However, the phenomenon of ice nucleation appeared within 20ns when switch the charge of iodine ions from -0.4e to 0.4e.

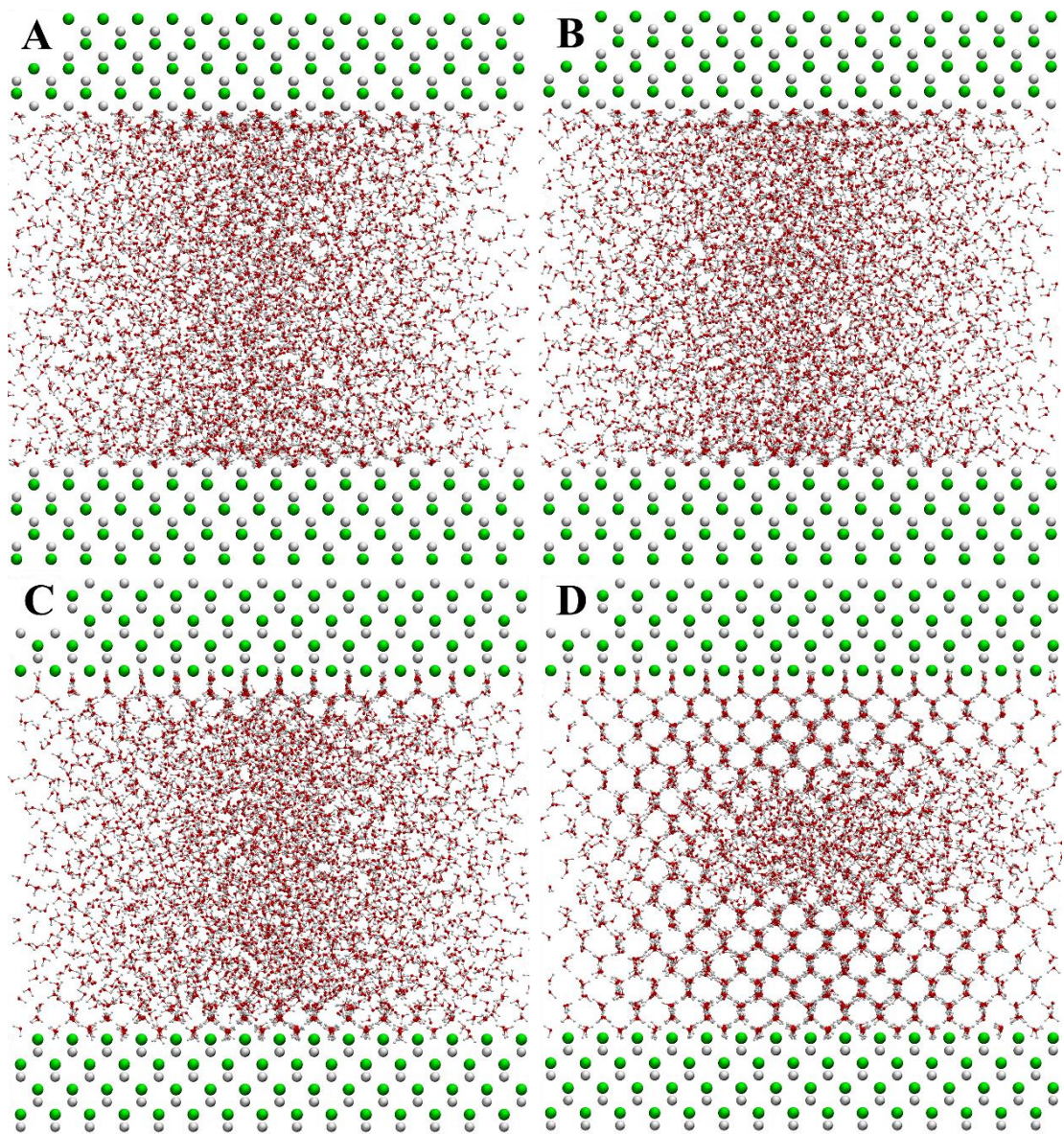

Figure 2: Snapshots of simulations on silver and iodine ions exposed $\gamma$-AgI (001) surface. Panels A and B are silver ions exposed as the outermost layers at 0ns, 200ns, respectively. Panels C and D are iodine ions exposed as the outermost layers at 0ns, 20ns, respectively. Silver ions, iodine ions, hydrogen atoms and oxygen atoms are silver, green, white and red, respectively. The snapshots are rotated to the direction of (110) plane.

According to the data from Table 1, ice nucleation is observed under specific conditions for dipoles on the surfaces of the substrates. Take silver ions exposed $\gamma$- AgI (001) surface as an example, the substrates with a surface charge between 0.2e and 0.6e can promotes ice formation, in constrast, the substrates with surface charges other than 0.2e to 0.6e can not promote ice formation. The behavior of water molecules at interfaces is central in understanding heterogeneous nucleation, therefore, it is crucial to analyze the behavior of water molecules at the interface in the simulation.

The microscopic structure of water molecules near the surface at the beginning of the simulation are focused. Density distribution of hydrogen atoms and oxygen atoms on silver and iodide exposed $\gamma$- AgI (001) surface at 0 ns are shown in Figure 3, and partial configurational snapshots of the water molecules at the interface are shown in Figure 3(a)-(f). For the silver exposed $\gamma$- AgI (001) surface, the charge of silver ions is 0.7e (Figure 3A), 0.4e (Figure 3B), -0.4e (Figure 3C). When the charge of ions exposed on the surface ranges from -1.0e to -0.1e, the state of water molecules at the interface are similar to Figure 3C, the charge of ions exposed on the surface ranges from 0.2e to 0.6e, the state of water molecules at the interface are similar to Figure 3B, the state of water molecules at the interface are similar to Figure 3A when the charge of ions exposed on the surface ranges from 0.7e to 1.0e. By analyzing the simulation trajectories, ice nucleation is observed in 15 ns when the charge of ions exposed on $\gamma$- AgI (001) surface equals to 0.4e, thus the density distribution curves of hydrogen atoms and oxygen atoms of silver exposed $\gamma$- AgI (001) surface at 15 ns are provided for comparison in Figure 3A-C, the density distribution curves of hydrogen atoms and oxygen atoms of iodide exposed $\gamma$- AgI (001) surface at 20 ns are provided for comparison in Figure 3D-3F.

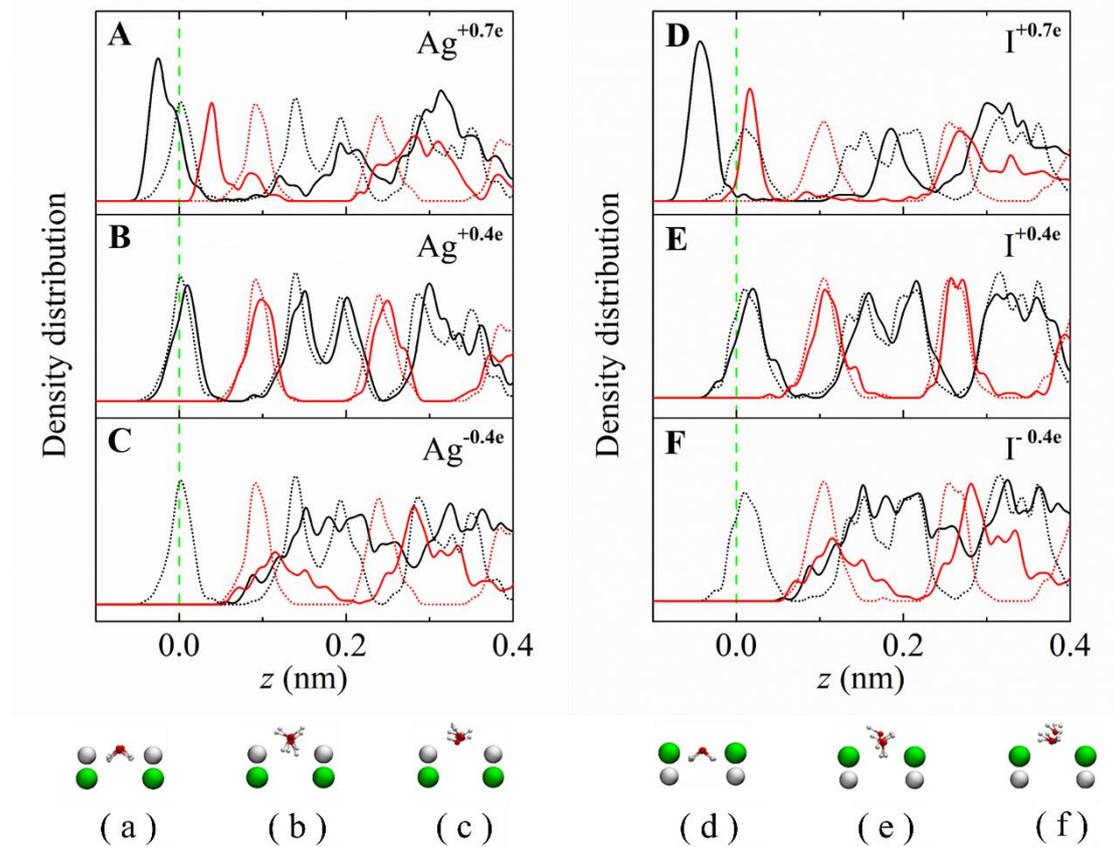

Figure 3: Density distribution profile of hydrogen atoms and oxygen atoms at 0 ns. the charge of panels A, B, C on silver exposed $\gamma$-AgI (001) surface are 0.7e, 0.4e, -0.4e, respectively, the charge of panels D, E, F on iodide exposed $\gamma$-AgI (001) surface are 0.7e, 0.4e, -0.4e, respectively. The curves of hydrogen atoms and oxygen atoms are black and red, respectively. The dot lines of hydrogen atoms and oxygen atoms at 15 ns with the charge of silver ions equals to 0.4e in panels A, B, C are black and red, respectively, and the dot lines of hydrogen atoms and oxygen atoms at 20 ns with the charge of iodine ions equals to 0.4e in panels D, E, F are black and red, respectively. panels (a)-(f) are snapshots of water near the surface Corresponded to the panels A-F. The vertical green thick dotted line coincides with the position of the center of the outermost ion of the surface. Silver ions, iodine ions, hydrogen atoms and oxygen atoms are silver, green, white and red, respectively.

Something interesting can be found after the comparison. In Figure 3B, density distribution curves of hydrogen atoms and oxygen atoms at 0ns are almost the same as the dotted lines at the interface, which means the system which the charge of silver ions equals to 0.4e in Figure 3B has generated the ice-like structure at 0ns, the peak of oxygen atoms generated is surrounded by the doublet of hydrogen atoms generated, and the truth is that this system nucleated ice within 15ns through observed the simulation trajectory of all atoms. As to Figure 3A and C, density distribution curves of hydrogen atoms and oxygen atoms do not match the dotted lines at all, that means these systems have not generated ice-like structure and no ice nucleation found in 200ns. For the iodide exposed $\gamma$- AgI (001) surface, the charge of iodine ions is 0.7e (Figure 3D), 0.4e (Figure 3E), -0.4e (Figure 3F). These curves showing the similar principles as silver exposed $\gamma$- AgI (001) surface.

It is worthy to remark that when the charge of ions exposed on $\gamma$- AgI (001) surface is 0.1e, the system can not nucleate ice. However, these systems are defined as incapable of nucleating ice does not mean that they will never nucleate ice,[4] a longer simulation time and a lower temperature may be helpful. But what can be determined is that these systems are not as effective as those systems which can nucleate ice in a shorter time.

Heterogeneous ice nucleation requires the formation of ice-like structure (Figure 3(b)) at the interface, and then other water molecules grow along this ice-like structure and into ice bulk finally.[9] Therefore, whether the substrate promotes the formation of ice-like structures is particularly important, thus the interaction between silver, iodide and hydrogen, oxygen should be understood.

Based on the above results, we recognize that when silver and iodine ions form a suitable dipole, it will guide water molecules near the substrates to adjust their patterns to form an ice-like structure, and then templated to generate bulk ice. In other words, silver iodide affects the patterns of the layer of water molecules near the substrate through dipole interaction to promote the ice nucleation. In order to verify this point, we selected a specific configuration as the initial configuration and reset the simulation parameters, this special configuration only formed a relatively ordered structure in the first layer near the substrate, and then transform the direction of the substrate's dipole to observe the simulation results. The specific operation method is an example of a system in which silver ions with a charge amount of 0.4e are exposed on the surface (Figure 1A-B). In this system, the substrate's ability in promoting ice nucleation has been proven (Table 1). The structure of 1.2ns (0.2ns after decrease the temperature, ice nucleatiion starts at 8ns) is focused after analyzing the density distribution curve of the trajectory, at this particular time, only a part of water molecules near the interface had formed a relatively ordered structure (Figure 3(b), 3(e)), and the water molecules in other places still remain in a state of chaos. the configuration at this time is therefore selected as the initial configuration. The difference from former(Figure 1A-B) is that we fix the water molecules in the first layer even some of them had not generated a relative ordered structure, and then change the dipole orientation of the substrate. The simulation time and the selection of other parameters has not changed unless otherwise specified.

In these simulations, the charge of silver ions gradually decreased from 0.4e to -0.4e. The simulation results showed that ice nucleation can be observed regardless of the charge of the silver ion. Figure 3 in Supporting Information provides a dynamic diagram to clearly observe the simulation results and fixed water molecules.

From the data in Table 1, it can be seen that the system can not freeze when the surface charge of the substrate is less than 0.2e. However, when the first layer of water molecules are fixed, ice nucleation can be observed whatever the charge of the silver ion is. The relationship between ice growth and time in simulations with different surface charges was shown in Supporting Information (Figure S4). We calculate the growth rate of ice molecules in different systems to analyze the ability of different substrates in promoting nucleation. Figure 4 shows that the growth rate of ice molecules increases with the increase of the surface charge, which shows that the different ability of ice nucleation in these systems.

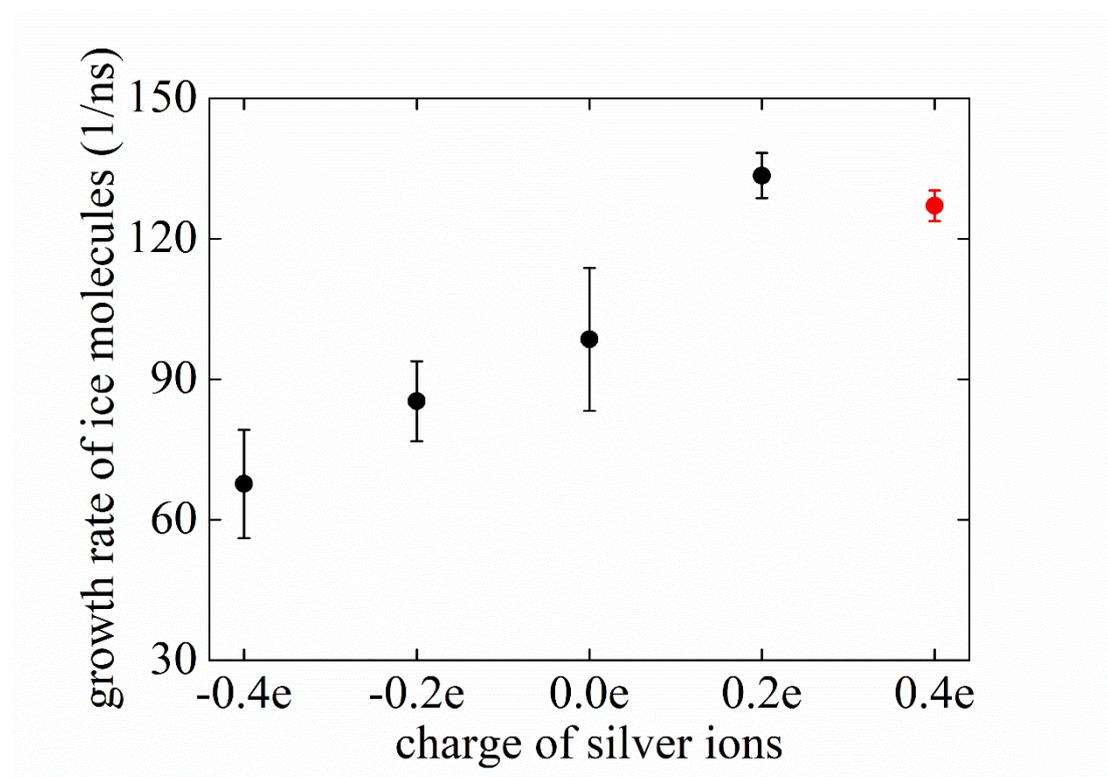

Figure 4: The relationship between the growth rate of ice molecules and the charge of silver ions. The first layer of water molecules near the interface in all systems are fixed, except that the substrate with a 0.4e charge of silver ions is used for comparison.

The silver iodide substrate begins to play a relatively secondary role after inducing the first layer of water molecules. If the charge of silver ions is between 0.2e-0.6e, it shows a promoting effect, otherwise it is a hindering effect, while the first layer of water molecules begins to play a relatively major role in promoting ice nucleation. It can be predicted that when the charge of silver ions equals to 0.4e, if the water molecules in the first layer are fixed, we can observe the phenomenon of ice nucleation in a shorter time.

When the charge range of silver ions on the substrate surface is between -0.4e and 0.4e, ice nucleation can be obserbed in these systems, which indicates that the fixed first layer of water molecules plays a leading role in the ice nucleation incident. The different charges of silver ions lead to different growth rates of ice molecules, which indicates that the dipole size of the substrate can delay or accelerate the ice nucleation phenomenon. However, when the charge of silver ions on the surface is between 0.2e and 0.4e, the substrates can promote the formation of a fine ice-like structure at the interface, so fixing the first layer of water molecules has no obvious effect on the freezing rate of water.

## Conclusion

We carry out several simulations of heterogeneous ice nucleation on silver and iodine ions exposed $\gamma$- AgI (001) surface, and change the charge of ions on the exposed surface, respectively. According to the simulations and analysis above, it is insufficient to predict the ice nucleation ability of silver iodide substrate even it has a possess a fine lattice match with ice. We analyzed the influence of the dipole of the silver iodide substrate on the water molecules at the interface and the influence of the water molecules at the interface on ice nucleation, which can be utilized in engineering to promote or inhibit ice nucleation. In the simulations we change the charge value of the silver iodide, but retain the lattice parameters. We remark that we do not aim at investigating specific systems like silver iodide surfaces in this study, but instead we intend to extract general principle and useful trends from idealized model substrates. Therefore, these conclusions can be applied to the substrate has a similar lattice parameter with silver iodide.

Supporting Information

Table S1

Table S1: Average freezing times at different charges on silver and iodine ions exposed γ-AgI (001) surface.

| Charge | 0.2e | 0.3e | 0.4e | 0.5e | 0.6e |
|---|---|---|---|---|---|
| Ag+ exposed | 19.755ns | 15.014 ns | 15.083 ns | 16.522 ns | 23.229 ns |
| I+ exposed | 99.376ns | 26.495 ns | 21.316ns | 61.559ns | - |

Figure S2

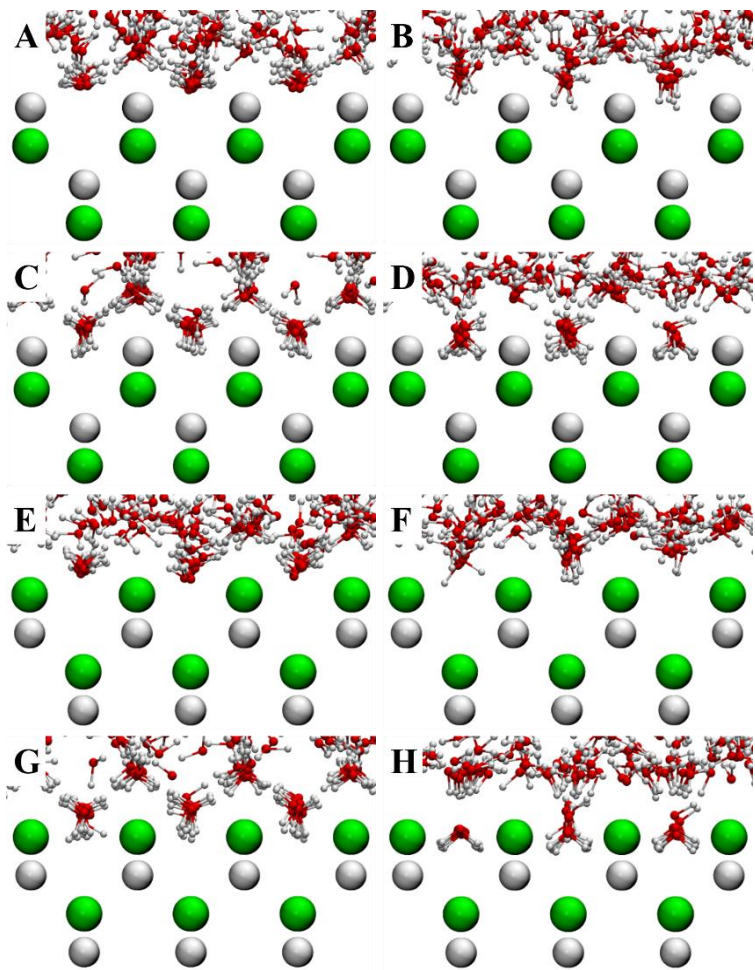

Figure S2: The position of water molecules near the exposed γ-AgI (001) surface at 1ns. The charge of panels A, B, C, and D on silver exposed γ-AgI (001) surface are -0.4e, 0.1e, 0.4e, and 0.7e, respectively, the charge of panels E, F, G, and H on iodide exposed $\gamma$-AgI (001) surface are -0.4e, 0.1e, 0.4e, and 0.7e, respectively. Silver ions, iodine ions, hydrogen atoms and oxygen atoms are silver, green, white and red, respectively.

Figure 3(dynamic picture)

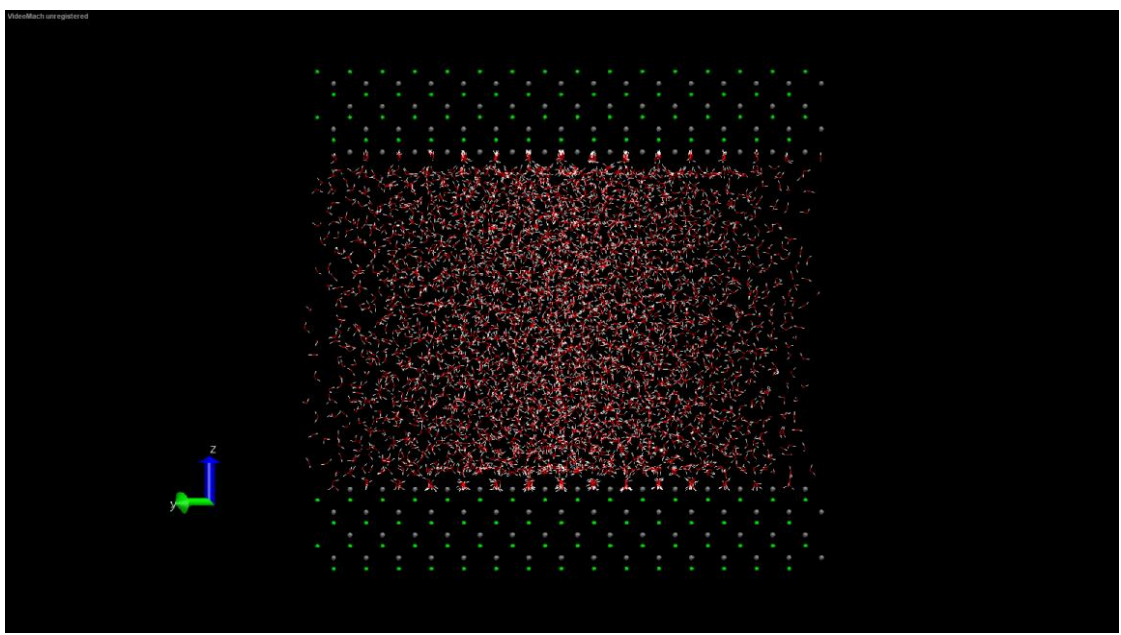

Figure S4

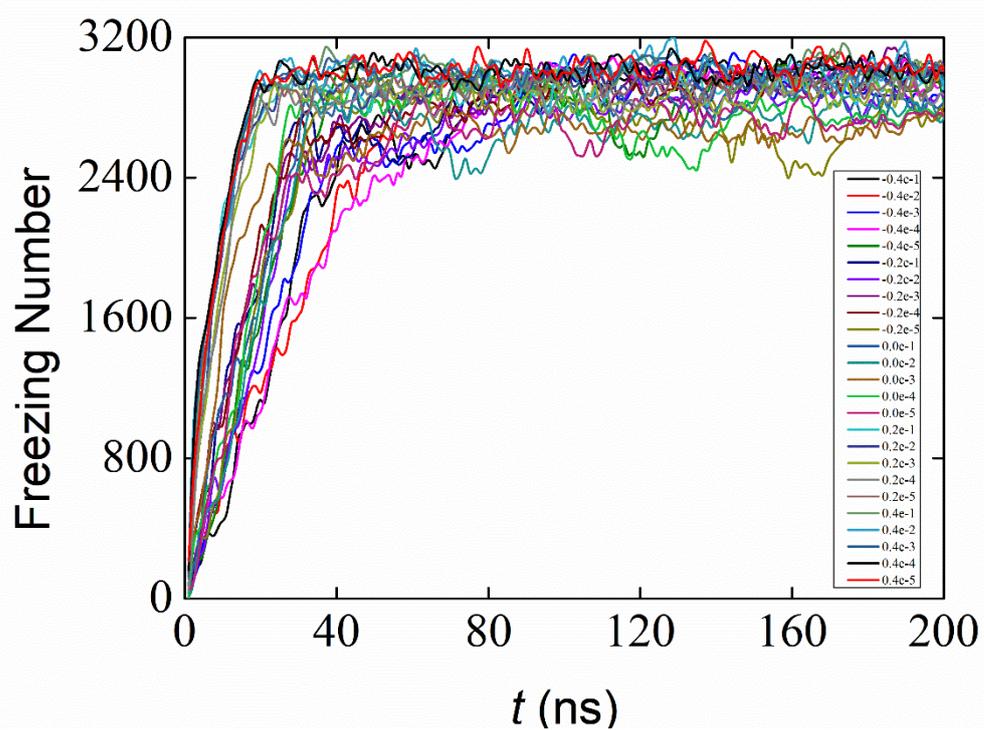

Figure S4: The relationship between ice growth and time in simulations with different surface charges. only part of the curves are shown for visual reasons.